\documentclass[prl,twocolumn,preprintnumbers,amsmath,amssymb,showpacs,
nofootinbib,floatfix]{revtex4}

\usepackage{graphicx,bm}

\makeatletter
\def\graphicscale{\twocolumn@sw{0.33}{0.4}}
\def\graphicthreescale{\twocolumn@sw{0.33}{0.4}}

\begin{document}

\title{Quasi-long-range order in trapped systems}

\author{Federico Crecchi}
\affiliation{Department of Physics, 
University of Chicago, 5720 S. Ellis Ave, Chicago, IL 60637, USA,} 
\affiliation{Scuola Normale Superiore, I-56126 Pisa, Italy} 
\author{Ettore Vicari}
\affiliation{Dipartimento di Fisica dell'Universit\`a di Pisa and
INFN, I-56127 Pisa, Italy} 
\date{November 20, 2010}

\begin{abstract}
  We investigate the effects of a trapping space-dependent potential
  on the low-temperature quasi-long-range order phase of
  two-dimensional particle systems with a relevant U(1) symmetry, such
  as quantum atomic gases. We characterize the universal features of
  the trap-size dependence using scaling arguments. The resulting
  scenario is supported by numerical Monte Carlo simulations of a
  classical two-dimensional XY model with a space-dependent hopping
  parameter whose inhomogeneity is analogous to that arising from the
  trapping potential in experiments of atomic gases.
\end{abstract}


\maketitle



Statistical systems are generally nonhomogeneous in nature, while homogeneous
systems are often an ideal limit of experimental conditions.  Thus, in the
study of critical phenomena, an important issue is how critical behaviors
develop in nonhomogenous systems.  Particularly interesting physical systems
are interacting particles constrained within a limited region of space by an
external potential.  This is a common feature of the experimental realizations
of the Bose-Einstein condensation (BEC) in diluted atomic vapors~\cite{CWK-02}
and of optical lattices of cold atoms~\cite{BDZ-08}, which have provided
a great opportunity to investigate the interplay between quantum and
statistical behaviors in particle systems.

In the BEC scenario, a macroscopic number of bosonic atoms accumulate
in a single quantum state and are described by a condensate wave
function, which naturally provides the complex order parameter
$\psi(x)$ of the phase transition and its relevant U(1) symmetry.
These global features characterize the XY universality class which
describes the universal critical behavior of a large class of systems,
see, e.g., Ref.~\cite{PV-02}.  The critical behavior arising from the
formation of the condensate in a trapped Bose gas has been
investigated experimentally~\cite{DRBOKS-07}, observing an increasing
correlation length compatible with the behavior expected at a
continuous transition of homogeneous systems belonging the
three-dimensional (3D) XY universality class.  Two-dimensional (2D)
homogeneous gases of bosonic particles do not show a real BEC with
decreasing the temperature $T$.  Neverthless, they are expected to
experience a finite-$T$ Kosterlitz-Thouless (KT)
transition~\cite{KT-73} separating the high-$T$ phase with
short-ranged correlations from a low-$T$ phase characterized by a
quasi-long-range order (QLRO), where the one-body correlation function
decays algebraically at large distance.  Experimental evidences of
such a transition in trapped Bose atomic gases have been provided in
Refs.~\cite{HSBBD-06,KHD-07,CRRHP-09}.

However, the inhomogeneity due to the trapping potential drastically
changes the general features of the critical behavior at the
transition separating the high-$T$ and low-$T$ phases, and, in the
case of 2D systems, of the QLRO phase. For example, correlation
functions of the critical modes do not develop a diverging length
scale in a trap.  The critical behavior of the unconfined homogeneous
system could be observed around the middle of the trap only when the
length scale $\xi$ of the correlations is much smaller than the length
scale $\xi_{t}$ induced by the trap size, and one looks at
small-distance correlations relatively to $\xi_t$.  If $\xi$ is large
but not much smaller than the trap size, the critical behavior gets
somehow distorted by the trap, although it gives rise to universal
effects in the large trap-size limit, controlled by the universality
class of the phase transition of the unconfined
system~\cite{CV-09,CV-10}.  The understanding of the trap effects is
necessary for an accurate determination of the critical parameters,
see, e.g., Refs.~\cite{CRRHP-09,PPS-10}.

In this paper we consider 2D systems showing a low-$T$ QLRO phase in
their phase diagram, after a KT transition.  We investigate how the
presence of the trap changes the main features of the QLRO of the
homogeneous system and, therefore, how one may get evidence of the
QLRO phase from the behavior of the system in the presence of the
trap.  For this purpose, we resort to a scaling analysis which allows
us to take into account the trap length scale when it becomes
sufficiently large, exploiting the universality of the scaling
behavior.

The above considerations also apply to other physically interesting
models, such as the Bose-Hubbard (BH) model~\cite{FWGF-89} at its
finite-$T$ superfluid transition, whose Hamiltonian in the presence of
confining potential reads
\begin{eqnarray}
{\cal H}_{\rm BH} &=& - {J\over 2}
\sum_{\langle ij\rangle} (b_i^\dagger b_j+ {\rm h.c.})
+ {U\over 2} \sum_i n_i(n_i-1) 
\nonumber \\
&& + \sum_i (\mu + v^2r^2) n_i,
\label{bhm}
\end{eqnarray}
where the sum runs over the bonds ${\langle ij \rangle }$ of a
$d$-dimensional lattice, $b_i$ are bosonic operators, $n_i\equiv
b_i^\dagger b_i$ is the particle density, $r$ the distance from the
center of the trap. The trap size is defined as $l \equiv {J^{1/2}/
v}$, see, e.g., \cite{CV-11}.  The BH model is of experimental
relevance because it describes cold bosonic atoms trapped in a limited
space region of optical lattices~\cite{JBCGZ-98}.  Other transitions
in the XY universality class are the $^4$He superfluid transition,
insulator-superconductor transitions, like that of the attractive
Hubbard model, etc....

In a standard scenario for a continuous transition, see, e.g.,
\cite{PV-02}, the critical behavior of a $d$-dimensional system is
characterized by two relevant parameters $u_t$ and $u_h$, which may be
associated with $T$, i.e., $u_t\sim T/T_c-1$ and the external field
$h$ coupled to the order parameter, with renormalization-group (RG)
dimension $y_t=1/\nu$ and $y_h= (d+2-\eta)/2$.  The presence of a trap
of size $l$ generally induces a further length scale $\xi_t$, which
must be taken into account to describe the critical
correlations. Within the trap-size scaling (TSS)
framework~\cite{CV-09,CV-10}, the scaling law of the singular part of
the free energy density around the center of the trap can be written
as
\begin{equation}
F_{\rm sing} = l^{-\theta d} {\cal F}(u_tl^{\theta y_t},
u_hl^{\theta y_h},xl^{-\theta})
\label{freee}
\end{equation}
where $\theta$ is the {\em trap exponent}.  
At the critical point ($u_t=0$), the length scale induced by the
trap behaves as $\xi_{t} \sim l^{\theta}$, and the correlation function of the
order parameter as
\begin{equation}
G(x,y)\equiv \langle \bar{\psi}(x) \psi(y) \rangle_c 
= l^{-\theta\eta} {\cal G}(xl^{-\theta},yl^{-\theta}). 
\label{twopf}
\end{equation}
Finite size effects, due to a finite volume
$L^d$, can be taken into account by adding a further dependence on $L
l^{-\theta}$ in the above scaling Ansatz~\cite{QSS-10}.

The value of $\theta$ depends on the way the external confining field
is coupled to the model variables.  In the case relevant for the
above-mentioned particle systems, the external trapping potential is
coupled to the particle density.
The corresponding perturbation can be inferred from the
many-body Hamiltonian~\cite{DGPS-99} in the presence of the external
potential $V(x)=v^p x^p$, i.e., $P_V=\int d^d x\, V(x) |\psi(x)|^2$,
where $\psi(x)$ is the complex order-parameter field.  The computation
of the RG dimensions of the trap parameter leads to~\cite{CV-09}
$\theta = p \nu/ (1+p\nu)$.  In order to apply it to the KT transition
of 2D U(1)-symmetric systems, we formally set $\nu=\infty$,
corresponding to the KT exponential behavior of the correlation length
$\xi \sim {\rm exp}(\tau^{-1/2})$ where $\tau \equiv
T/T_{c}-1\rightarrow 0^+$, thus obtaining $\theta=1$ for any power $p$
of the potential.  For comparison, we mention that $\theta=0.57327(4)$
at the 3D BEC transition in a harmonic trap~\cite{CV-09}. Moreover, in
a Gaussian theory perturbed by $P_V$, since $\nu=1/2$ we have
$\theta=p/(2+p)$, for any spatial dimension.  It is worth mentioning
that analogous TSS behaviors~\cite{CV-10}, with the same trap
exponents, apply to the quantum $T=0$ superfluid-Mott transition of
the BH model (\ref{bhm}) at fixed integer density, which belongs to
the ($d$+1)-dimensional XY universality class~\cite{FWGF-89}.

A standard representative model of the 2D XY universality class is the
classical square-lattice XY model,
\begin{equation}
H = - J \sum_{\langle ij \rangle } {\rm Re}
\, \bar{\psi}_i \psi_j, \quad \psi_i\equiv e^{i\varphi_i}\in {\rm U}(1), 
\label{XYmodel}
\end{equation}
which presents the same universal features of 2D systems
whose phase diagram shows a KT transition between high-$T$ and
low-$T$ QLRO phases.  We may further exploit universality to
investigate the effects of an inhomogeneity analogous to that of 2D
trapped particle systems, where the external confining potential is
generally coupled to the particle density, which can be associated
with an energy-density operator in a corresponding effective model.  A
2D XY model with an external space-dependent field coupled to the
energy density is obtained by considering a space-dependent hopping
parameter $U_{ij}$,~\cite{hoptrap}
\begin{eqnarray}
&& H_U = -  J \sum_{\langle ij \rangle } 
 {\rm Re} \, \bar{\psi}_i U_{ij} \psi_j,
\label{xymodtr} \\
&& U_{ij} = 1 + V(r_{ij}),\quad V(r) = v^p r^p,
\label{hoppingt}
\end{eqnarray}
where $p$ is an even positive integer, $r_{ij}$ is the distance from
the origin of the midpoint of nearest-neighbor sites. We set $J=1$.
The inhomogeneity arising from the space dependence of $U_{ij}$ is
analogous to that arising from a trapping potential in particle
systems, such as the BH model (\ref{bhm}).  Thus, $l\equiv 1/v$ may be
considered as the analog of the trap size. At large distance, since
$V(r)\to\infty$, the spin variables get effectively frozen.  When
$p\to \infty$ the effect of the external potential $V$ is equivalent
to confining a homogeneous system in a box of size $L=l=1/v$, and the
TSS becomes the standard finite-size scaling (FSS).  At the critical
KT temperature $T_c=0.893(1)$~\cite{HP-97} of the homogeneous model
(\ref{XYmodel}), the TSS of the model (\ref{xymodtr}) is expected to
follow the scaling Ansatz (\ref{freee}) and (\ref{twopf}), with
$\eta=1/4$ and $\theta=1$ as computed above by RG arguments.

We now turn to the QLRO phase, which is the main issue of this paper.  The
homogeneous model is critical in the whole low-$T$ region $T<T_c$, where the
correlation function $\langle \bar{\psi}_x \psi_y \rangle$ decays as
$1/|x-y|^{\eta(T)}$ with a $T$-dependent exponent $\eta(T)$:
$\eta(T)=T/(2\pi)+O(T^2)$, increasing up to $\eta(T_c)=1/4$ (some numerical estimates 
are reported in Refs.~\cite{APV-08,Berche-03}).  This critical
behavior is controlled by a line of Gaussian fixed points, essentially given
by the spin-wave theory $H_{\rm sw}=\int d^2x \,(\nabla \varphi)^2$, which is
the leading nontrivial term for $T\to 0$.  We may apply the same spin-wave
approximation to infer the value of $\theta$ controlling the TSS in the QLRO
phase.  In the spin-wave limit we obtain
\begin{equation}
H_{\rm sw} = \int d^2 x \, {1\over 2} (1 + v^p r^p) (\nabla \varphi)^2.
\label{swham}
\end{equation}
The trap exponent $\theta$ is related to the RG dimension $y_v$ of parameter
$v$, $\theta=1/y_v$, which can be obtained from the relation $py_v-p +
y_{(\partial_\mu\varphi)^2}=d$ taking also into account that
$y_{(\partial_\mu\varphi)^2}=d$ in the spin-wave theory. We eventually obtain
$\theta=1$ independently of the power $p$.

Therefore, we have $\theta=1$ at very low-$T$, where the spin-wave
approximation holds, and at the KT transition at $T=T_c$.  A natural
scenario is that the TSS has universal features in the whole QLRO,
with $\theta=1$ for any $T\le T_c$.  We shall provide
numerical evidence of this scenario.
  
\begin{figure}[tbp]
\includegraphics*[scale=\graphicscale]{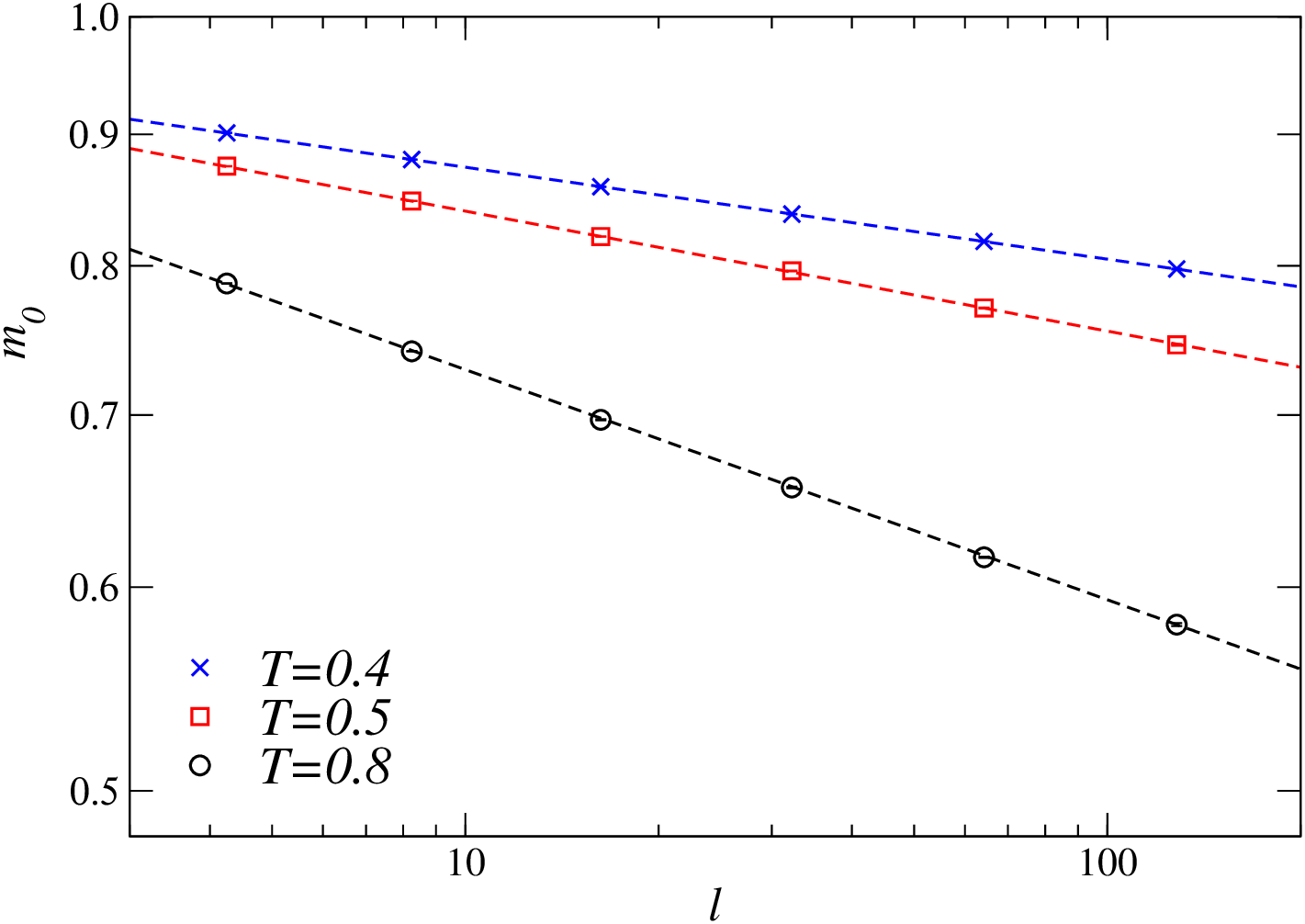}
\caption{Log-log plot of $m_0\equiv \langle \psi_0\rangle$
for the model (\ref{xymodtr}),
  at $T=0.4,\,0.5,\,0.8$, for $L/l\simeq 2$~\cite{MCdetails}.  Statistical
  errors are hardly visible.  The lines show fits to $a l^{-\zeta/2}$.  }
\label{m0}
\end{figure}

\begin{figure}[tbp]
\includegraphics*[scale=\graphicscale]{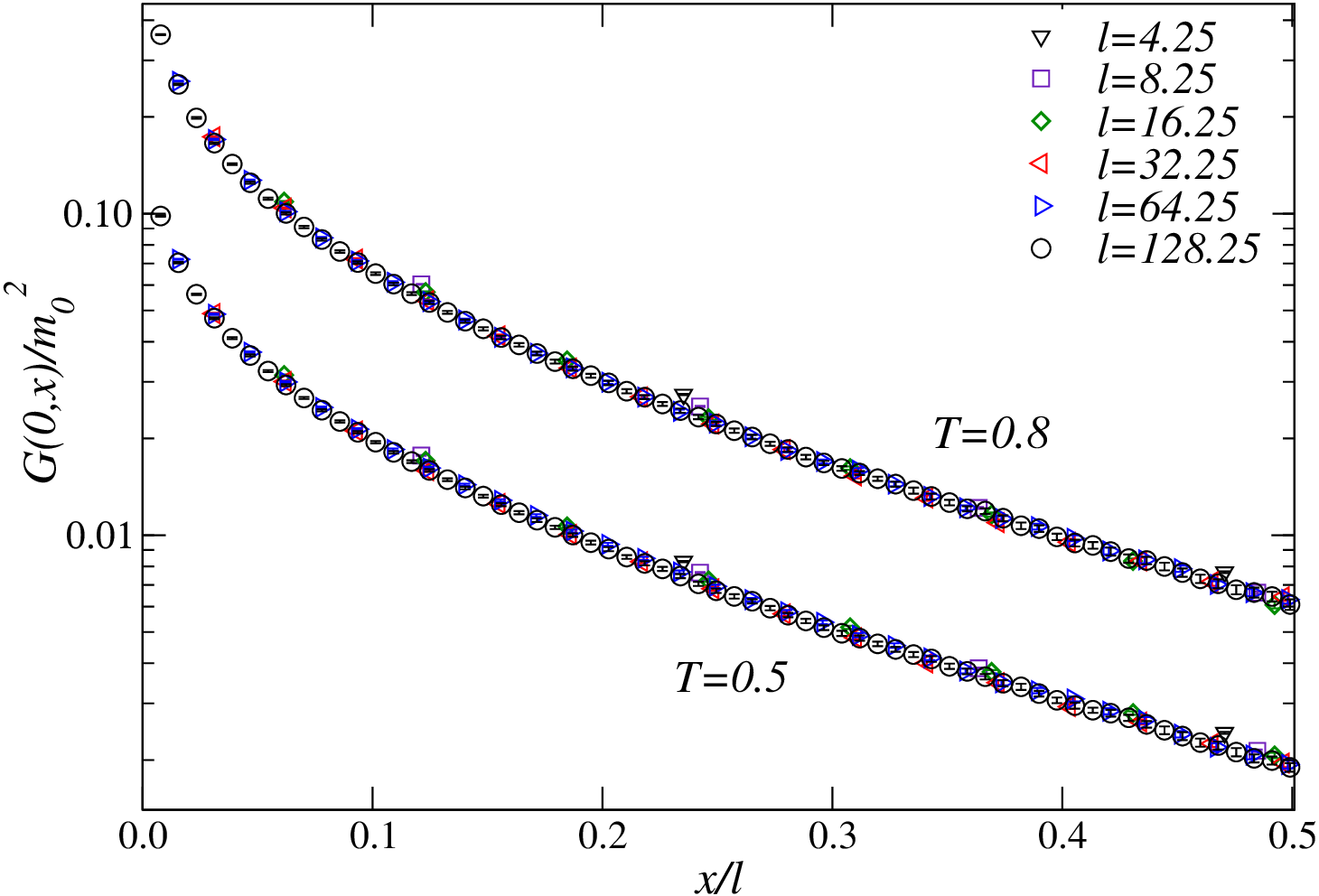}
\caption{$G(0,x)/m_0^2$ vs $x/l$ for the model (\ref{xymodtr}),
at $T=0.5,\,0.8$,
  for several values of $l$, and $L/l\simeq 2$~\cite{MCdetails}.  
The sets of data at fixed $T$ are clearly
  converging to a nontrivial large-$l$ limit }
\label{gtT}
\end{figure}

For this purpose, we present numerical results for the model
(\ref{xymodtr}) with $V(r)=(r/l)^2$, for $T<T_c\simeq 0.893$, obtained
by Monte Carlo (MC) simulations.  We center the {\rm trap} in the
middle of a square lattice $(2L+1)\times(2L+1)$.  The parameter $L$
and $l$ are chosen so that the spin variables close to the boundaries
are effectively frozen, making unnecessary the use of larger lattices.
We use fixed boundary conditions $\psi_{b}=1$.~\cite{MCdetails}

We consider the local magnetization at the origin and the two-point
correlation function, defined as
\begin{equation}
m_0 \equiv \langle \psi_{0} \rangle,
\quad 
G(\vec{x},\vec{y}) \equiv \langle \bar{\psi}_{\vec{x}} \psi_{\vec{y}} 
\rangle -  \langle \bar{\psi}_{\vec{x}} \rangle \langle 
\psi_{\vec{y}} \rangle .
\label{maco}
\end{equation}
The boundary conditions $\phi_b=1$ breaks the U(1) symmetry, thus
allowing a nonzero local magnetization.  According to the above scaling
considerations, we expect that their asymptotic trap-size dependence is
\begin{eqnarray}
&&m_0 \sim  l^{-\eta(T)\theta/2}, 
\label{moscal} \\
&&G(\vec{x},\vec{y}) \approx
l^{-\eta(T)\theta} {\cal G}(\vec{x}/l^\theta,\vec{y}/l^\theta),
\label{gscal}
\end{eqnarray}
where $\theta=1$ and $\eta(T)$ is the $T$-dependent exponent
of the homogeneous system.  Since $\theta=1$, a nontrivial simultaneous FSS
and TSS limit can be achieved by keeping $L/l$ fixed, where scaling behaviors
analogous to Eqs.~(\ref{moscal}) and (\ref{gscal}) 
apply.~\cite{scacorr}

Fig.~\ref{m0} shows data of $m_0$ for some values of $T<T_c$.  A power-law
trap-size dependence is clearly supported by the data.  Fits to $a
l^{-\zeta/2}$, see Fig.~\ref{m0}, give $\zeta=0.072(1),\,0.093(1),\,0.179(1)$
respectively for $T=0.4,\,0.5,\,0.8$.  Since according to Eq.~(\ref{moscal})
$\zeta=\eta\theta$ and $\theta=1$, these results should be compared with the
available estimates of $\eta(T)$ obtained in homogeneous systems, which are in
good agreement~\cite{estimates}.  

Fig.~\ref{gtT} shows results for
$G(0,\vec{x})$.  In agreement with Eqs.~(\ref{moscal}) and (\ref{gscal}), they
show that $G(0,\vec{x})/m_0^2 = g(x/l)$ in the large-$l$ limit. At small
distance $x\ll l$, $g(y)\sim y^{-\eta(T)}$, to recover the behavior of the
homogeneous system.  

The spatial dependence of the local
magnetization does not show a simple scaling behavior.  The numerical results
appear consistent with
\begin{equation}
\langle \psi_x \rangle \approx l^{-\eta(T_X)/2} f(X)
\label{mx}
\end{equation}
where $X\equiv x/l$ and $T_X\equiv T/(1+X^2)$, which can be derived using
arguments based on a local-temperature approximation, noting that $T_X$ may be
considered as an effective local (space-dependent) temperature.

\begin{figure}[tbp]
\includegraphics*[scale=\graphicscale]{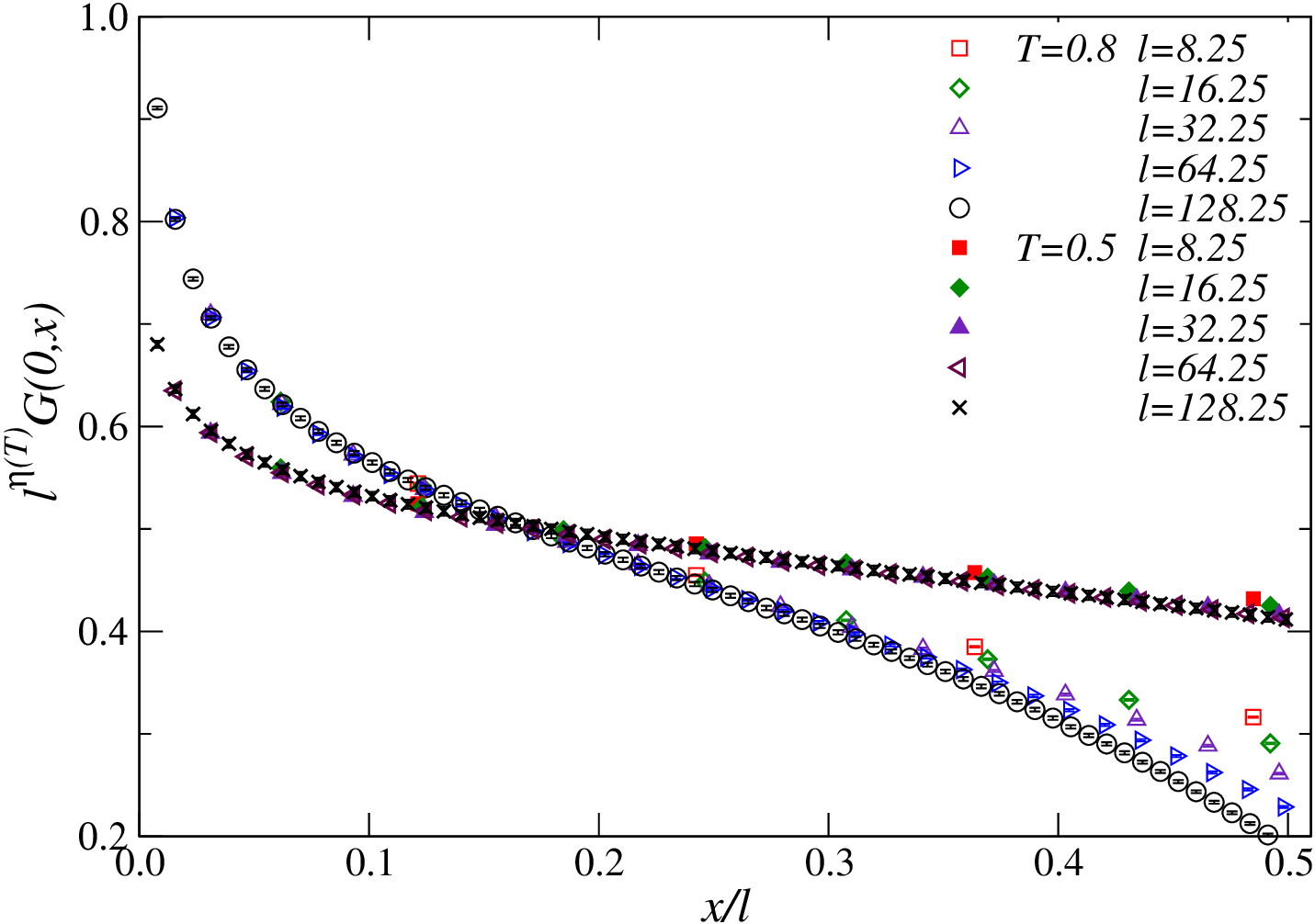}
\caption{$l^{\eta(T)}G(0,x)$ vs $x/l$,
for the $U_{ij}$ given in Eq.~(\ref{hopping2}),
at $T=0.5,\,0.8$,
  for several values of $l$, and $L/l\approx 2$.
}
\label{gtT2nd}
\end{figure}

We also consider a 2D XY model where the hopping parameter decreases
moving far from the origin, i.e., replacing
\begin{equation}
U_{ij} = [1 + V(r_{ij})]^{-1}, \quad V(r)=v^2r^2,
\label{hopping2}
\end{equation}
in Eq.~(\ref{xymodtr}).  In this case the regions far from the origin
are effectively in the high-$T$ phase.  The lattice system is set as
before, but we use open boundary conditions which are compatible with
a diverging hopping parameter at large distance.  Thus, $\langle
\psi_x \rangle =0$ everywhere, including the origin.  MC simulations
show that the TSS is again characterized by the trap exponent
$\theta=1$. This is shown by Fig.~\ref{gtT2nd}, where we report
$l^{\eta(T)}G(0,x)$ for several values of $l$, $L/l\approx
2$~\cite{fsseff}, with $\eta(T)$ obtained from the data of
Fig.~\ref{m0}.  The sets of data for $T=0.5$ and $T=0.8$ are clearly
converging to a nontrivial large-$l$ limit, at least up to $X\equiv
x/l$ corresponding to $T_X\simeq T/(1+X^2) = T_c\approx 0.893$
($X_c\approx 0.89,\, 0.34$ for $T=0.5,\,0.8$
respectively).~\cite{nota}

In conclusion, we have characterized the trap-size dependence within
the low-$T$ QLRO phase and at the KT finite-$T$ transition of 2D
trapped systems.  Using scaling arguments, we have argued that it is
described by the TSS Ansatz (\ref{moscal}) and (\ref{gscal}) with the
trap exponent $\theta=1$ in the whole QLRO phase, up to the KT
transition.  This scenario has been supported by numerical results for
classical 2D XY models with space-dependent hopping parameters, which
give rise to inhomogeneities analogous to that of trapped atomic gases
in actual experiments.  These results should be useful to get evidence
of QLRO in trapped systems, and also determine the critical parameters
at the KT transition, such as $T_c$ in model (\ref{xymodtr}).  For
example, the KT critical point corresponds to a TSS with $\eta=1/4$,
while values $\eta<1/4$ corresponds to the low-$T$ QLRO phase.


\begin{thebibliography}{99}
\bibitem{hoptrap} As put forward in Ref.~\cite{RBSMJ-10}, a trap for atomic
  systems in an optical lattice, described by the BH model, may be realized
  through space-dependent hopping parameters.
\bibitem{scacorr}
Scaling corrections
are suppressed by powers of $l$. Since $\theta=1$, they should be
$O(l^{4-1/\eta(T)})$ as in the FSS of the homogeneous system~\cite{HPV-05}.
\bibitem{estimates} Estimates of $\eta$ for the square-lattice
  nearest-neighbor XY model at $T=0.4,\,0.5,\,0.8$ are respectively
  $\eta=0.079(6),\,0.098(7),\,0.171(3)$ from Ref.~\cite{APV-08}, and
  $\eta=0.074(6),\,0.100(8),\,0.19(2)$ from Ref.~\cite{Berche-03}.
\bibitem{fsseff} Finite-$L$ effects turn out to be stronger in this case.  The
  data shown in Fig.~\ref{gtT2nd} were obtained using
  $c_v=4$~\cite{MCdetails}, corresponding to $L/l\approx 2$.  We also
  performed MC simulations with $c_v=16$, thus $L/l\simeq 4$, which showed a
  scaling behavior consistent with Eq.~(\ref{gscal}) as well.
\bibitem{nota} The behavior for $X>X_c$ is not completely clear.  Our
 MC data, in particular for $L/l\approx 4$ up to $L=256$ (not shown
 here), suggest a vanishing large-$l$ limit for $X>X_c$.
\end{thebibliography}
\end{document}